%% file: pepm.tex
\documentclass[11pt]{article} 
\usepackage{fullpage,url}
\newenvironment{descit}[1]{\begin{quote} \textit{#1}}{\end{quote}}

\input{psfig-dvips}

\newif\ifpdf
\ifx\pdfoutput\undefined
  \pdffalse
\else
  \pdfoutput=1
  \pdftrue
\fi

\ifpdf
  \usepackage[pdftex]{graphicx}
  \usepackage[pdftex]{color}
  \DeclareGraphicsExtensions{.pdf,.png,.jpg}
\else
  \usepackage[dvips]{graphicx}
  \usepackage[dvips]{color}
  \DeclareGraphicsExtensions{.eps,.epsi,.ps}
\fi

\usepackage{times}

\def\midv{\mathop{\,|\,}}

\long\def\cbk#1{{\color{red}[CBK: #1]}}
\newlength\colwidth 

\title{Mixed-Initiative Interaction = Mixed Computation\footnote{This
work is supported in part by US National
Science Foundation grants DGE-9553458 and IIS-9876167.}}

\author{Naren Ramakrishnan, Robert Capra, and Manuel A. P\'{e}rez-Qui\~{n}ones\\ 
Department of Computer Science\\
Virginia Tech, Blacksburg, VA 24061, USA\\
Contact Email: {\tt naren@cs.vt.edu}}

\begin{document}

\maketitle

\begin{abstract}
\noindent
We show that partial evaluation can be usefully viewed as 
a programming model for realizing mixed-initiative 
functionality in interactive applications. 
Mixed-initiative interaction between two participants is one 
where the parties can take turns at any time to change 
and steer the flow of interaction. We concentrate on 
the facet of mixed-initiative referred to as `unsolicited 
reporting' and demonstrate how out-of-turn interactions
by users can be modeled by `jumping ahead' to nested 
dialogs (via partial evaluation).  Our approach permits 
the view of dialog management systems in terms of their 
native support for staging and simplifying interactions; 
we characterize three different voice-based interaction 
technologies using this viewpoint. In particular, we 
show that the built-in form interpretation algorithm (FIA) 
in the VoiceXML dialog management architecture is actually 
a (well disguised) combination of an interpreter 
and a partial evaluator.
\end{abstract}

\newpage
\section{Introduction} 
\label{intro}
Mixed-initiative interaction~\cite{computational-mixed} has been studied 
for the past 30 years in the areas of artificial intelligence 
planning~\cite{prodigy}, human-computer interaction~\cite{mixed-hci}, and 
discourse analysis~\cite{coulthard}. As Novick and Sutton point 
out~\cite{mixed-notkin}, 
it is `one of those things that people think that they can recognize
when they see it even if they can't define it.' It
can be broadly viewed as a flexible interaction strategy between 
participants where the parties can take turns at any time to change 
and steer the flow of interaction. Human conversations are typically
mixed-initiative and, interestingly, so are interactions with some modern
computer systems. Consider the following two dialogs with a 
telephone pizza delivery service that has voice-recognition
capability (the line numbers are provided for ease of reference):

\begin{descit}{Dialog 1}
\vspace{-0.1in}
\begin{tabbing}
[x] \= abcdefab \= thiscanactuallybeamuchlongersentenceokay \kill
0 \> {\bf Caller:} \> $\prec$calls Joe's Pizza on the phone$\succ$ \\
1 \> {\bf System:} \> Thank you for calling Joe's pizza ordering system.\\
2 \> {\bf System:} \> What size pizza would you like?\\
3 \> {\bf Caller:} \> I'd like a medium, please.\\
4 \> {\bf System:} \> What topping would you like on your pizza?\\
5 \> {\bf Caller:} \> Pepperoni.\\
6 \> {\bf System:} \> What type of crust do you want?\\
7 \> {\bf Caller:} \> Uh, deep-dish.\\
8 \> {\bf System:} \> So that is a medium pepperoni pizza with deep-dish crust.
             Is this correct?\\
9 \> {\bf Caller:} \> Yes.\\
(conversation continues to get delivery and payment information)
\end{tabbing}
\end{descit}

\begin{descit}{Dialog 2}
\vspace{-0.1in}
\begin{tabbing}
[x] \= abcdefab \= thiscanactuallybeamuchlongersentenceokay \kill
0 \> {\bf Caller:} \> $\prec$calls Joe's Pizza on the phone$\succ$ \\
1 \> {\bf System:} \> Thank you for calling Joe's pizza ordering system.\\
2 \> {\bf System:} \> What size pizza would you like?\\
3 \> {\bf Caller:} \>  I'd like a sausage pizza, please.\\
4 \> {\bf System:} \> Okay, sausage.\\
5 \> {\bf System:} \> What size pizza would you like?\\
6 \> {\bf Caller:} \> Medium.\\
7 \> {\bf System:} \> What type of crust do you want?\\
8 \> {\bf Caller:} \> Deep-dish.\\
9 \> {\bf System:} \>  So that is a medium sausage pizza with deep-dish crust.
             Is this correct?\\
10 \> {\bf Caller:} \> Yes.\\
(conversation continues to get delivery and payment information)
\end{tabbing}
\end{descit}

\noindent
Both these conversations involve the specification of a (size,topping,crust)
tuple to complete the pizza ordering procedure but differ 
in important ways. In the first dialog, the caller responds to the 
questions in the order they are posed by the system. The system has the
initiative at all times (other than, perhaps, on line 0) and such an 
interaction is thus
referred to as {\it system-initiated}. In the second dialog, when the
system prompts the caller about pizza size, he responds
with information about his choice of topping instead
(sausage; see line 3 of {\it Dialog 2}). Nevertheless, the conversation
is not stalled and the system continues with the other aspects of the
information gathering activity. In particular, the system registers that the
caller has specified a topping, skips its default question on this topic,
and repeats its question about the size (see line 5 of {\it Dialog 2}). The 
caller 
thus gained the initiative for a brief period during the conversation, 
before returning it to the system. A conversation that `mixes' these modes
of interaction in such arbitrary ways is said to be {\it mixed-initiative}.

\subsection{Tiers of Mixed-Initiative Interaction}
\label{tiers}
It is well accepted that mixed-initiative provides a more natural and
personalized mode of interaction. A matter of debate, however, are
the qualities that an interaction must possess to merit its
classification as mixed-initiative~\cite{mixed-notkin}. In fact,
determining who has the initiative at a given point in an interaction can itself 
be a contentious issue! The role of intention in
an interaction and the underlying task goals also affect the characterization
of initiative. We will not attempt to settle this debate here but 
a few preliminary observations will be useful.

One of the basic levels of mixed-initiative is referred to
as {\it unsolicited reporting} in~\cite{allen-intelligent} and is illustrated
in {\it Dialog 2} above. In this facet, a participant 
provides information out-of-turn (in our case the caller, about his
choice of topping). Furthermore, the out-of-turn
interaction is not agreed upon in advance by the two participants. 
Novick and Sutton~\cite{mixed-notkin} stress that the unanticipated 
nature of out-of-turn interactions is important and that mere turn-taking 
(perhaps in a hardwired order) does not constitute 
mixed-initiative. Finally, notice that in {\it Dialog 2} there is a resumption
of the original questioning task once processing of the unsolicited response
is completed. In other applications, an unsolicited response might shift the
control to a new interaction sequence and/or abort the current interaction.

Another level of mixed-initiative involves {\it subdialog invocation};
for instance, the computer system might not have understood the user's
response and ask for clarifications (which amounts to it having 
the initiative). A final, sophisticated, form of mixed-initiative is one 
where participants negotiate with each other to determine initiative 
(as opposed to merely `taking the initiative')~\cite{allen-intelligent}:

\vspace{-0.05in}
\begin{descit}{Dialog 3}
\vspace{-0.1in}
\begin{tabbing}
abcdefabyr \= thiscanactuallybeamuchlongersentenceokay \kill
(with apologies to O. Henry)\\
{\bf Husband:} \> Della, Something interesting happened today that I want to
tell you.\\
{\bf Wife:} \> I too have something exciting to tell you, Jim.\\
{\bf Husband:} \> Do you want to go first or shall I tell you my story?
\end{tabbing}
\end{descit}

In addition to models that characterize initiative, there are models
for designing dialog-based interaction systems.
Allen et al.~\cite{allen-ai} provide a taxonomy of such software models
--- finite-state machines, slot-and-filler
structures, frame-based methods, and more sophisticated models involving
planning, agent-based programming, and exploiting contextual information.
While mixed-initiative interaction can be studied in any of these models,
it is beyond the scope of this paper to address all or even a majority
of them. 

Instead, we concentrate on the view of (i) a dialog as a
task-oriented information assessment activity requiring the filling of a
set of slots,
(ii) where one of the participants in the dialog is a computer 
system and the other 
is a human, and (iii) where mixed-initiative arises from unsolicited 
reporting (by the human), involving out-of-turn 
interactions. This characterization includes many voice-based
interfaces to information (our pizza ordering dialog is an example) and
web sites modeling interaction by hyperlinks~\cite{pipe-tois}. 
In Section~\ref{ourmodel}, we show that partial evaluation can be 
usefully viewed as a programming model for such applications. 
Section~\ref{voice-tech} presents three different voice-based interaction
technologies and analyzes them in terms of their native support for
mixing initiative. Finally, Section~\ref{future}
discusses other facets of mixed-initiative and mentions
other software models to which our approach can be extended.

\vspace{-0.1in}
\section{Programming a Mixed-Initiative Application}
\vspace{-0.03in}
\label{ourmodel}
Before we outline the design of a system such as Joe's Pizza, we introduce
a notation~\cite{levinson,goffman} that captures basic elements 
of initiative and response in an interaction sequence. The notation expresses
the local organization of a dialog~\cite{manuel-thesis,manuel-chi} as 
adjacency pairs; for instance, {\it Dialog 1} is represented as:

\vspace{-0.05in}
{\center
\begin{tabbing}
(Ic \= Rs) \= (Is \= Rc) \= (Is \= Rc) \= (Is \= Rc) \= (Is \= Rc) \kill
(Ic \> Rs) \> (Is \> Rc) \> (Is \> Rc) \> (Is \> Rc) \> (Is \> Rc) \\
\,\,0 \> \,\,1 \> \,\,2 \> \,\,3 \> \,\,4 \> \,\,5 \> \,\,6 \> \,\,7 \> \,\,8 \> \,\,9 \\
\end{tabbing}}

\noindent
The line numbers given below the interaction sequence refer to the utterance
numbers in the dialog presented in Section~\ref{intro}. 
The letter I denotes who has the initiative --- caller (c) or the system (s) ---
and the letter R denotes who provides the response. It is easy to see
from this notation that {\it Dialog 1}  consists
of five turns and that the system has the initiative for the last
four turns.  The initial turn is modeled as the caller having the
initiative because he or she chose to place the phone call in the first place.
The system quickly takes the initiative after playing a greeting to
the caller (which is modeled here as the response to the caller's call). 
The subsequent four interactions then address three questions and a 
confirmation, all involving the system retaining the initiative (Is) and
the caller in the responding mode (Rc). Likewise,
the mixed-initiative interaction in {\it Dialog 2} is 
represented as:

{\center
\begin{tabbing}
(Ic \= Rs) \= (Iso \= (Ic \= Rs) \= Rc) \= (Is \= Rc) \= (Is \= Rc) \kill
(Ic \> Rs) \> (Is \> (Ic \> Rs) \> Rc) \> (Is \> Rc) \> (Is \> Rc) \\
\,\,0 \> \,\,1 \> \,\,2,5 \> \,\,3 \> \,\,4 \> \,\,6 \> \,\,7 \> \,\,8 \> \,\,9 \,\,10 \\
\end{tabbing}}

\noindent
In this case, the system takes the initiative in utterance 2 but instead
of responding to the question of size in utterance 3, the caller 
takes the initiative, causing
an `insertion' to occur in the interaction sequence (dialog)~\cite{levinson}. 
The system responds with an acknowledgement (`Okay, sausage.') in
utterance 4. This is represented as the nested pair (Ic Rs) above.
The system then re-focuses the dialog on the question of pizza size in
utterance 5 (thus retaking the initiative). In utterance 6 the
caller responds with the desired size (medium) and
the interaction proceeds as before, from this point. 

The notation is useful to describe the space of possible interactions that
are to be supported. For instance, utterances 0 and 1 have to proceed in order.
Utterances dealing with selection of (size,topping,crust) can then
be nested in any order and provide interesting
opportunities for mixing initiative. 
For instance, if a user is a frequent
customer of Joe's Pizza, he might take the initiative and specify all three
pizza attributes on the first available prompt:

\begin{descit}{Dialog 4}
\vspace{-0.1in}
\begin{tabbing}
[x] \= abcdefab \= thiscanactuallybeamuchlongersentenceokay \kill
0 \> {\bf Caller:} \> $\prec$calls Joe's Pizza on the phone$\succ$ \\
1 \> {\bf System:} \> Thank you for calling Joe's pizza ordering system.\\
2 \> {\bf System:} \> What size pizza would you like?\\
3 \> {\bf Caller:} \>  I'd like a sausage pizza, medium, and deep-dish.\\
(conversation continues with confirmation of order) 
\end{tabbing}
\end{descit}

\noindent
Finally, the utterances dealing with confirmation of the user's request 
can proceed only after choices of all three pizza attributes have been 
made. There are 13 possible interaction sequences (discounting permutations
of attributes specified in a given utterance) --- 1 possibility
of specifying everything in one utterance, 6 possibilities of specification
in two utterances, and 6 possibilities of specification in three
utterances. If we include permutations, there are 24 possibilities (our
calculations do not consider situations where, for instance, the system doesn't 
recognize the user's input and reprompts for information).

\begin{figure}
\centering
\begin{tabular}{cc}
\includegraphics[height=1.95in]{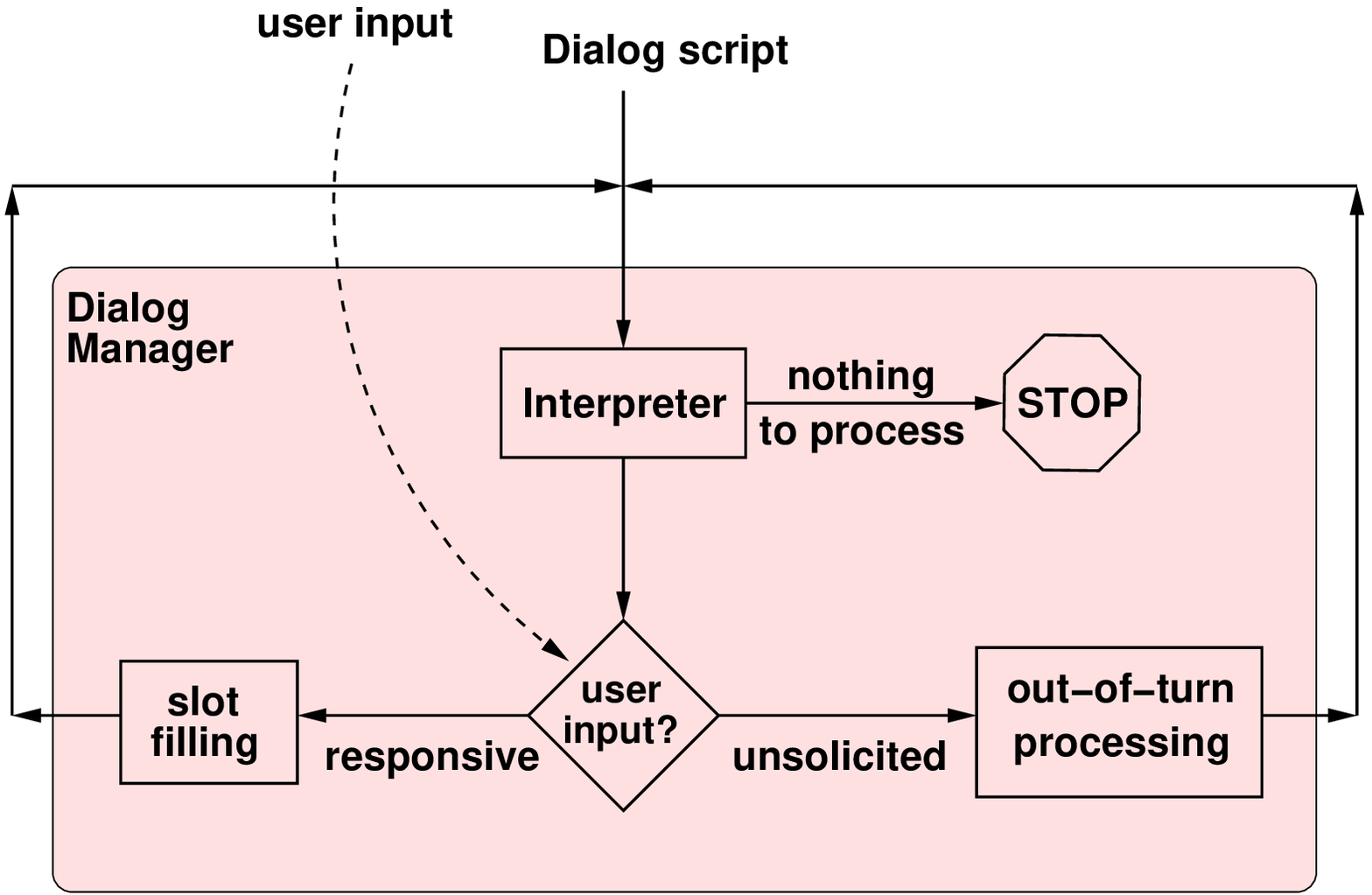}
\hspace{0.1in}
\includegraphics[height=1.95in]{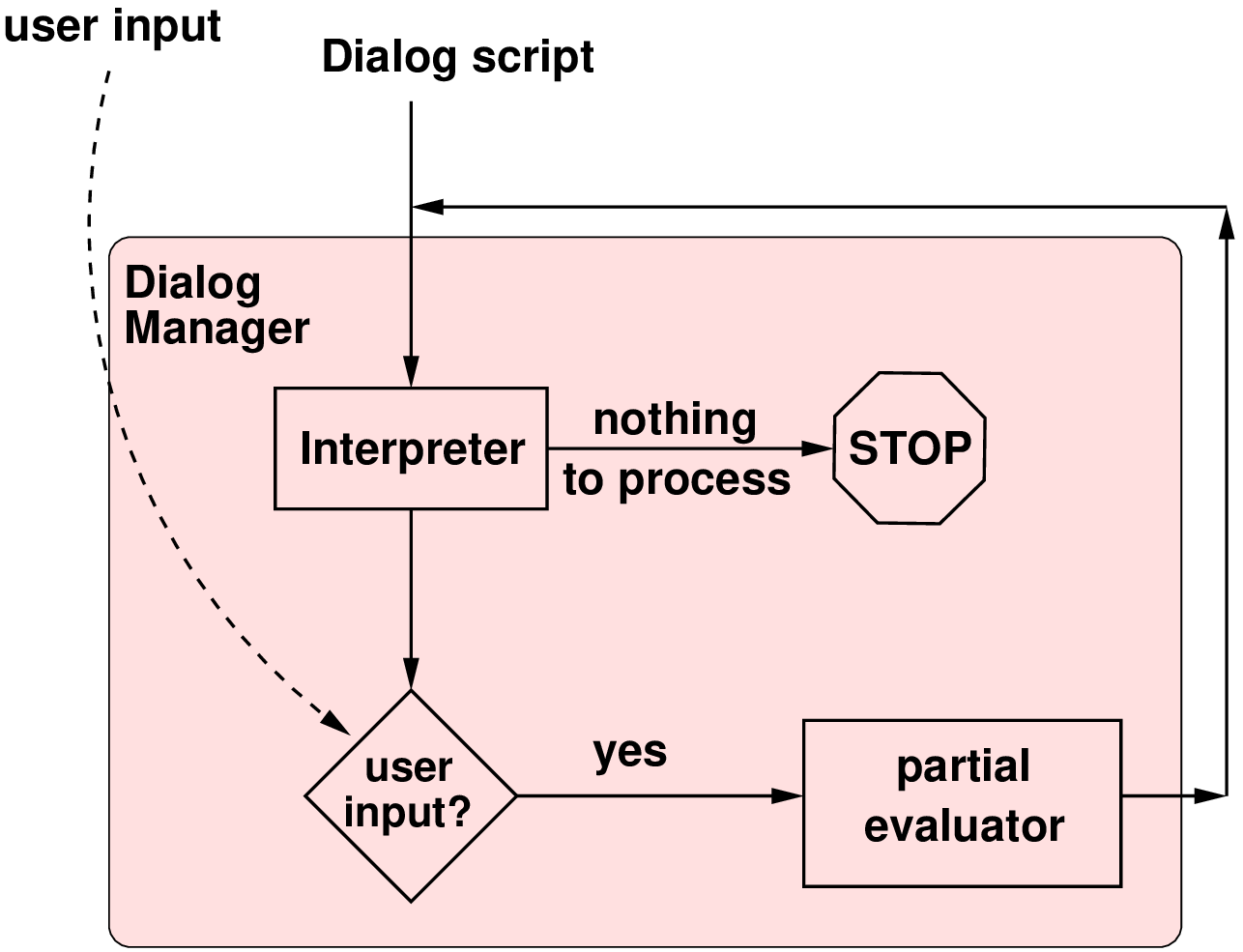}
\end{tabular}
\caption{Designs of software systems for 
mixed-initiative interaction. (left) Traditional system architecture,
distinguishing between responsive and unsolicited inputs.
(right) Using partial evaluation to handle inputs uniformly.}
\label{dialog-designs}
\end{figure}

Many programming models 
view mixed-initiative sequences as requiring some
special attention to be accommodated. In particular, they rely on recognizing 
when a user has provided unsolicited 
input\footnote{We use the term `unsolicited
input' here to refer to expected but out-of-turn inputs as opposed to
completely unexpected (or out-of-vocabulary) inputs.}
and qualify a 
shift-in-initiative as a `transfer of 
control.'
This implies that the mechanisms that handle out-of-turn interactions 
are often different 
from those that realize purely system-directed interactions. 
Fig.~\ref{dialog-designs} (left) describes a typical software design. 
A dialog manager is in charge of
prompting the user for input, queueing messages onto an output 
medium, event processing, and managing the overall flow of interaction.
One of its inputs is a dialog script that contains a specification of
interaction and a set of slots that are to be filled. In our pizza example,
slots correspond to placeholders for values of size, topping, 
and crust. An interpreter determines the first unfilled
slot to be visited and presents any prompts 
for soliciting user input.
A responsive input requires mere slot filling whereas unsolicited inputs 
would require out-of-turn processing (involving a combination of slot 
filling and simplification). In turn, this causes a revision of the 
dialog script. The interpreter terminates when there is nothing left to
process in the script. While typical dialog managers
perform miscellaneous functions such as error control,
transferring to other scripts, and accessing scripts from a database, the 
architecture in Fig.~\ref{dialog-designs}
(left) focuses on the aspects most relevant to our presentation.

Our approach, on the other hand, is to think of a mixed-initiative
dialog as a program,
all of whose arguments are passed by reference and which correspond to the
slots comprising information assessment. As usual, an interpreter
in the dialog manager
queues up any applicable prompts to an output medium.
Both responsive and
unsolicited inputs by a user now correspond (uniformly)
to values for arguments; they are processed by partially evaluating 
the program with respect to these inputs. The result of partial evaluation
is another dialog (simplified as a result of user input) which is handed
back to the interpreter. This novel design
is depicted in Fig.~\ref{dialog-designs} (right) and a dialog script 
represented in a programmatic notation is given in Fig.~\ref{pizza-script}. 
Partial evaluation of Fig.~\ref{pizza-script} with respect to user
input will remove the conditionals for all slots that
are filled by the utterance (global variables are assumed to be
under the purview of the interpreter).
The reader can verify that a sequence of such partial evaluations
will indeed mimic the interaction sequence depicted in {\it Dialog 2} 
(and any of the other mixed-initiative sequences).

Partial evaluation serves two critical uses in our design. The first is
obvious, namely the processing of out-of-turn interactions (and any
appropriate simplifications to the dialog script). The more subtle advantage
of partial evaluation is its support for staging mixed-initiative
interactions. The 
mix-equation~\cite{jones,jones-pe-book} holds for every possible way
of splitting inputs into two categories, without enumerating and
`trapping' the ways
in which the computations can be staged. 
For instance, the nested pair in {\it Dialog 2} is supported as a natural 
consequence of our design, not by anticipating and reacting to an 
out-of-turn input. 

Another way to characterize the system designs in Fig.~\ref{dialog-designs} is
to say that Fig.~\ref{dialog-designs} (left) makes a distinction of responsive
versus unsolicited inputs, whereas Fig.~\ref{dialog-designs} (right) makes
a more fundamental
distinction of fixed-initiative (interpretation) versus
mixed-initiative (partial
evaluation). In other words, Fig.~\ref{dialog-designs} (right) carves
up an interaction sequence into (i) turns that are 
to be handled in the order they are modeled (by an interpreter),
and (ii) turns that can involve mixing of initiative (handled
by a partial evaluator). 
In the latter case, the computations are actually used as 
a {\it representation of interactions.} Since only mixed-initiative 
interactions involve multiple staging options
and since these are handled by the partial
evaluator, our design requires the {\it least} amount of specification 
(to support all interaction sequences). For instance, the script
in Fig.~\ref{pizza-script} models the parts that involve mixing of
initiative and helps realize all of the 13 possible interaction sequences.
At the same time it does not model the confirmation sequence of 
{\it Dialog 2} because the caller cannot confirm his order before selecting 
the three pizza attributes! This turn should be handled by 
straightforward interpretation. 

\begin{figure}
\centering
\begin{tabular}{|l|} \hline
{\tt pizzaorder(size,topping,crust)}\\
\{\\
\,\,\,\, {\tt if (unfilled(size))\{}\\
\,\,\,\,\,\,\,\, {\tt /* prompt for size */}\\
\,\,\,\, {\tt \}}\\
\,\,\,\, {\tt if (unfilled(topping))\{}\\
\,\,\,\,\,\,\,\, {\tt /* prompt for topping */}\\
\,\,\,\, {\tt \}}\\
\,\,\,\, {\tt if (unfilled(crust))\{}\\
\,\,\,\,\,\,\,\, {\tt /* prompt for crust */}\\
\,\,\,\, {\tt \}}\\
\}\\
\hline
\end{tabular}
\caption{Modeling a dialog script as a program parameterized by slot variables
that are passed by reference.}
\label{pizza-script}
\end{figure}

To the best of our knowledge, such a model of partial evaluation
for mixed-initiative interaction 
has not been proposed before. An extensive literature
search has revealed no related prior work. 
While computational models for mixed-initiative
interaction remain an active area of research~\cite{computational-mixed}, 
such work is characterized by keywords such as `controlling mixed-initiative
interaction,' `knowledge representation and reasoning strategies,' and
`multi-agent co-ordination.' There are even projects that talk about
`integrating' mixed-initiative interaction and partial evaluation to realize
an architecture for planning and learning~\cite{prodigy}. We are optimistic
that our work has the same historical significance as the relation
between explanation-based generalization (a learning technique in AI) 
and partial evaluation established by van Haremelen and Bundy 
in 1988~\cite{EBG_PE}.

\vspace{-0.1in}
\section{Software Technologies for Voice-Based Mixed-Initiative Applications}
\label{voice-tech}
One of the main contributions of our model is that it characterizes the
minimum amount of information needed to program a mixed-initiative interaction
sequence.
Once a programmer supplies a script such 
as Fig.~\ref{pizza-script} mixed-initiative 
interaction is obtained, quite literally, `for free.' This means that 
we can use the design in Fig.~\ref{dialog-designs} (right) as a benchmark 
to compare and contrast the amount of specification required in other 
approaches. 

As indicated in Section~\ref{tiers}, our model is applicable to 
voice-based interaction technologies as well as web access via hyperlinks.
We concentrate on voice-based applications since interaction with web
sites is addressed in a related paper~\cite{pipe-tois} and because 
the design constraints in voice-based applications pose interesting
considerations for our model. In addition, a variety of commercial
technologies are available for voice-based applications (in contrast to
web sites) that will aid in comparative assessment.

\begin{figure}
\centering
\begin{tabular}{cc}
\includegraphics[height=2in]{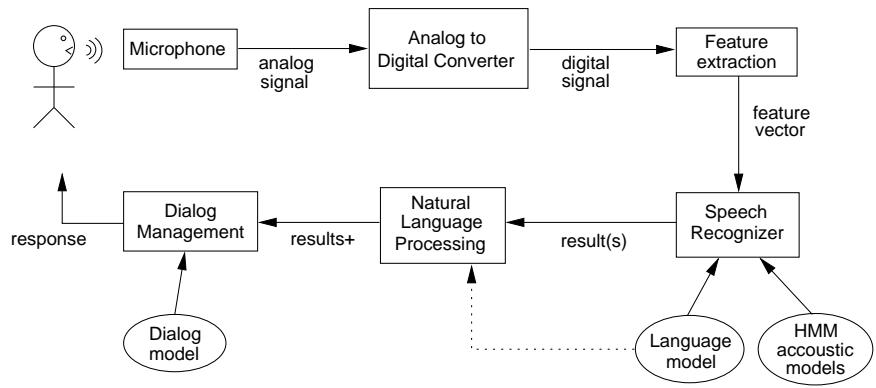}
\end{tabular}
\caption{Basic components of a spoken language processing system.}
\label{spreco}
\end{figure}

\vspace{-0.1in}
\subsection{Basic Principles of Voice-Based Interaction}
Before we can study the programming of mixed-initiative in
a voice-based application, it will be helpful to understand
the basic architecture (see Fig.~\ref{spreco})
of a spoken language processing system. As a user speaks into the 
system, the sounds produced are 
captured by a microphone and converted into a digital signal by an 
analog-to-digital converter. In telephone-based systems
(the VoiceXML architecture covered later in the paper is geared toward
this mode), the microphone is part of the
telephone handset and the analog-to-digital conversion is typically done by
equipment in the telephone network (in some cellular telephony models,
the conversion would be performed in the handset itself).

The next stage (feature extraction) prepares the digital speech signal to be
processed by the speech recognizer. Features of the signal important for 
speech recognition are extracted from the original signal, organized as an
attribute vector, and passed to the recognizer.  

Most modern speech recognizers use Hidden Markov Models (HMMs) and associated
algorithms to represent, train, and recognize speech.  HMMs are
probabilistic models that must be trained on an input set of data.  A common
technique is to create sets of acoustic HMMs that model phonetic units of
speech in context. These models are created from a training set of speech
data that is (hopefully) representative of the population of users who will
use the system. A language model is also created prior to performing
recognition. The language model is typically used to specify valid
combinations of the HMMs
at a word- or sentence-level.  In this way, the
language model specifies the words, phrases, and sentences that the recognizer
can attempt to recognize.  The process of recognizing a new input speech
signal is then accomplished using efficient search algorithms (such as Viterbi
decoding) to find the best matching HMMs, given the constraints of the language
model.  The output of the speech recognizer can take several different forms,
but the basic result is a text string that is the recognizer's best guess of
what the user said.  Many recognizers can provide additional information such
as a lattice of results, or an N-best ranked list of results (in case the later
stages of processing wish to reject the recognizer's top choice).  A good
introduction to speech recognition is available in~\cite{martin-speech}.

The stages after speech recognition vary depending on the application and the 
types of processing required. Fig.~\ref{spreco} presents
two additional phases that are commonly included in spoken language processing 
systems in one form or another. We will broadly refer to the first 
post-recognition stage as natural language processing (NLP). NLP is a 
large field in its own right and includes many sub-areas such as parsing, 
semantic interpretation, knowledge representation, and speech acts; an 
excellent introduction is available in Allen's classic~\cite{allen-nlp}. Our 
presentation in this paper has assumed NLP support for slot-filling (i.e.,
determining values for slot variables from user input).

This is commonly achieved by defining parts of a language model
and associating them with slots. The language model could take two 
major forms --- context-free grammars and statistical-based (such 
as n-grams). Here we
focus on the former: in this approach, slots can be specified within
the productions of a context-free grammar (akin to a attribute
grammar) or they can be associated with
the non-terminals in the grammar.

We will refer to the next phase of processing as simply `dialog management'
(see Fig.~\ref{spreco}). In this phase, augmented results from 
the NLP stage are incorporated into the dialog and any associated logic
of the application is executed. The job of the dialog manager is to 
track the proceedings of the dialog and to generate appropriate 
responses. This is often done within some logical processing 
framework and a dialog model (in our case, a dialog script)
is supplied as input that is specific to the particular application being 
designed. The execution of the logic on the dialog model (script) results 
in a response that can be presented back to the user. Sometimes
response generation is separated out into a subsequent stage.

\begin{figure}
\centering
\begin{tabular}{cc}
\includegraphics[width=3in]{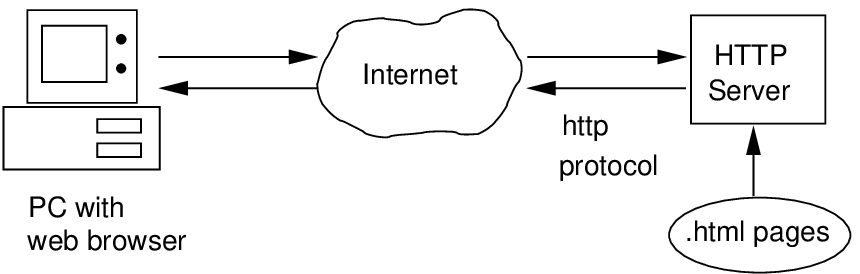}
\hspace{0.1in}
\includegraphics[width=3.9in]{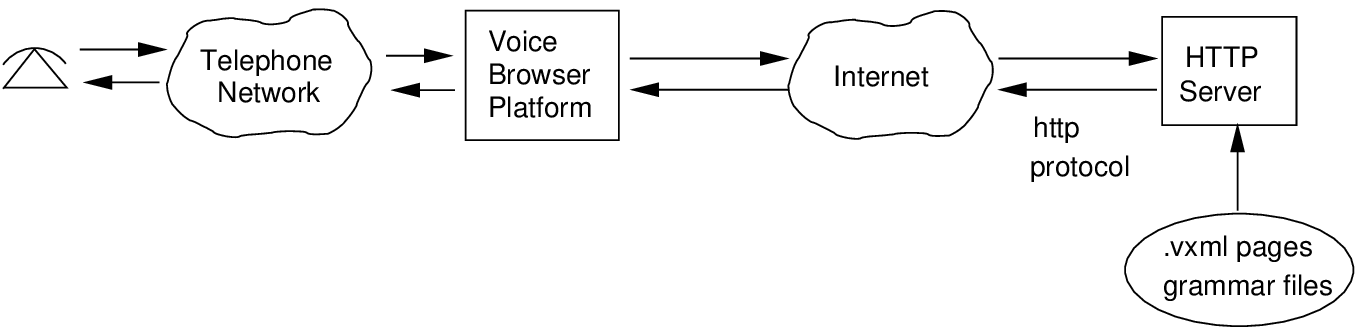}
\end{tabular}
\caption{(left) Accessing HTML documents via a HTTP web server.
(right) Accessing VoiceXML documents via a HTTP web server.}
\label{html-vxml}
\end{figure}

\begin{figure}
\centering
\small
\begin{verbatim}
<?xml version="1.0"?>
<vxml version="1.0">
<!-- pizza.vxml
     A simple pizza ordering demo to illustrate some basic elements
     of VoiceXML.  Several details have been omitted from this demo
     to help make the basic ideas stand out. -->
  <form id="welcome">
    <block name="block1">
      <prompt> Thank you for calling Joe's pizza ordering system. </prompt>
      <goto next="#place_order" />
    </block>
  </form>

  <form id="place_order">
    <field name="size">
      <prompt> What size pizza would you like? </prompt>
    </field>

    <field name="topping">
      <prompt> What topping would you like on your pizza? </prompt>
    </field>

    <field name="crust">
      <prompt> What type of crust do you want? </prompt>
    </field>

    <field name="verify">
      <prompt>
        So that is a <value expr="size"/> <value expr="topping"/> pizza
        with <value expr="crust"/> crust.
        Is this correct?
      </prompt>
      <grammar> yes | no </grammar>
    </field>

    <filled>
      <if cond="verify=='no'">
         <clear namelist="size topping verify crust"/>
         <prompt> Sorry.  Your order has been canceled. </prompt>
      <else/>
         <prompt>Thank you for ordering from Joe's pizza.</prompt>
      </if>
    </filled>

  </form>
</vxml>
\end{verbatim}
\caption{Modeling the pizza ordering dialog in a VoiceXML document.}
\label{vpizza}
\end{figure}

\vspace{-0.1in}
\subsection{The VoiceXML Dialog Management Architecture}
There are many technologies and delivery mechanisms available for
implementing Fig.~\ref{spreco}'s basic components. A popular 
implementation can be seen in the VoiceXML dialog management architecture.
VoiceXML is a markup language designed to simplify the construction of
voice-response applications~\cite{voicexml}. Initiated by a 
committee comprising AT\&T, IBM, Lucent Technologies, and Motorola,
it has emerged as a standard in telephone-based voice user interfaces 
and in delivering web content via voice. We will hence cover this 
architecture in detail.

The basic idea is to describe interaction
sequences using a markup notation in a VoiceXML {\it document.} As the
VoiceXML specification~\cite{voicexml} indicates, a VoiceXML document
constitutes a conversational finite state machine and describes a
sequence of interactions (both fixed- and mixed-initiative are supported).
A web server can serve VoiceXML documents using the HTTP 
protocol (Fig.~\ref{html-vxml} (right)), just as easily as HTML documents 
are currently served over the Internet (Fig.~\ref{html-vxml} (left)).
In addition, voice-based applications require a suitable delivery
platform, illustrated by a telephone in Fig.~\ref{html-vxml} (right). The 
voice-browser platform in Fig.~\ref{html-vxml} (right)
includes the VoiceXML interpreter which processes the
documents, monitors user inputs, streams messages,
and performs other functions expected of a dialog management system. Besides
the VoiceXML interpreter, the voice-browser platform includes speech 
recognizers, speech synthesizers, and telephony interfaces to help 
realize important aspects of voice-based interaction.

Dialog specification in a VoiceXML document involves organizing
a sequence of {\it forms} and {\it menus}. Forms specify a set of
slots (called field item variables) that are to be filled by user input. Menus
are syntactic shorthands (much like a {\tt case} construct); since they
involve only one field item variable (argument), there are no opportunities
for mixing initiative. We do not discuss menus further in this paper.
An example VoiceXML document for our pizza application is given 
in Fig.~\ref{vpizza}.

\begin{figure}
\small
\centering
\begin {tabular}{|p{6.6in}|}\hline
\vspace{-0.2in}
\begin{verbatim}
#JSGF V1.0;

grammar sizetoppingcrust;

public <sizetoppingcrust> =
   <size> {this.size=$} [<topping> {this.topping=$}] [<crust> {this.crust=$}] |
   <size> {this.size=$} <crust> {this.crust=$} <topping> {this.topping=$} |
   <topping> {this.topping=$} [<crust> {this.crust=$}] [<size> {this.size=$}] |
   <topping> {this.topping=$} <size> {this.size=$} <crust> {this.crust=$} |
   <crust> {this.crust=$} [<size> {this.size=$}] [<topping> {this.topping=$}] |
   <crust> {this.crust=$} <topping> {this.topping=$} <size> {this.size=$};

<size> = small | medium | large;
<topping> =  sausage | pepperoni | onions | green peppers;
<crust> = regular | deep dish | thin;
\end{verbatim}
\\\hline
\end{tabular}
\caption{A form-level grammar to be used in conjunction with
the script in Fig.~\ref{vpizza} to realize mixed-initiative interaction.
The above productions for {\tt sizetoppingcrust} cover all possibilities
of filling slot variables from user input, including multiple slots filled
by a given utterance, and various permutations of specifying pizza attributes.}
\label{formgram}
\end{figure}

As shown in Fig.~\ref{vpizza}, the pizza dialog consists of two forms. The
first form ({\tt welcome}) merely welcomes the user and transitions to the 
second. The {\tt place\_order} form involves four {\tt field}s 
(slot variables) --- the first three cover the pizza attributes and the
fourth models the confirmation variable (recall the dialogs in
Section~\ref{intro}). In particular, prompts for soliciting user input
in each of the fields are specified in Fig.~\ref{vpizza}. 

Interactions in a VoiceXML application proceed just like a web application
except instead of clicking on a hyperlink (to goto a new state), the
user talks into a microphone. The VoiceXML interpreter then
determines the next state to transition to. Any appropriate responses
(to user input) and prompts are
delivered over a speaker. The 
core of the interpreter is a so-called form interpretation algorithm 
(FIA) that drives the interaction. 
In Fig.~\ref{vpizza}, the fields provide for a fixed-initiative, system-directed
interaction. The FIA simply visits all fields in the order they are presented
in the document.
Once all fields are filled, a check is made to ensure that
the confirmation was successful; if not, the fields are cleared (notice
the {\tt clear namelist} tag) and the FIA will proceed to 
{\tt prompt} for the inputs again,
starting from the first unfilled field --- {\tt size}. 

\begin{figure}
\centering
\small
\begin {tabular}{|p{6in}|}\hline
\vspace{-0.2in}
\begin{verbatim}
While (true)
{
        // SELECT PHASE
        Select the first form item with an unsatisfied guard condition
             (e.g., unfilled)
          If no such form item, exit
	
        // COLLECT PHASE
        Queue up any prompts for the form item 
        Get an utterance from the user
		
        // PROCESS PHASE
        foreach (slot in user's utterance)
        {
               if (slot corresponds to a field item) {
                    copy slot values into field item variables
                    set field item's `just_filled' flag
               }
        }
        // some code for executing any `filled' actions triggered
}
\end{verbatim}
\\\hline
\end{tabular}
\caption{Outline of the form interpretation algorithm (FIA) in
the VoiceXML dialog management architecture. Adapted from~\cite{voicexml}.}
\label{fiaalgo}
\end{figure}

\begin{figure}
\small
\centering
\begin {tabular}{|p{5in}|}\hline
\vspace{-0.2in}
\begin{verbatim}
#JSGF V1.0;

grammar sizetoppingcrust;

public <sizetoppingcrust> = word*;

word = <size> {this.size=$} | 
       <crust> {this.crust=$} | 
       <topping> {this.topping=$};

<size> = small | medium | large;
<topping> =  sausage | pepperoni | onions | green peppers;
<crust> = regular | deep dish | thin;
\end{verbatim}
\\\hline
\end{tabular}
\caption{A alternative form-level grammar to realize mixed-initiative
interaction with
the script in Fig.~\ref{vpizza}.}
\label{formgram2}
\end{figure}

The form in Fig.~\ref{vpizza} is referred to as a directed one since the
computer has the initiative at all times and the {\tt field}s are filled
in a strictly sequential order. To make the interaction mixed-initiative
(with respect to {\tt size}, {\tt crust}, and {\tt topping}),
the programmer merely has to specify a so-called
{\it form-level grammar} that describes possibilities for slot-filling from
a user utterance. An example
form-level grammar file ({\tt sizetoppingcrust.gram}) that covers all 
possibilities is given in
Fig.~\ref{formgram}. The grammar is associated with the dialog script
by including the line:
\begin{verbatim}
    <grammar src="sizetoppingcrust.gram" type="application/x-jsgf"/>
\end{verbatim}
just before the definition of
the first {\tt field} (size) in Fig.~\ref{vpizza}.

The form-level grammar contains productions for the various choices 
available for size, topping, and crust and also qualifies 
all possible parses for
a given utterance (modeled by the non-terminal {\tt sizetoppingcrust}). Any
valid combination of the three pizza aspects uttered by the user (in
any order) is recognized and the appropriate slot variables are instantiated.
To see why this also achieves mixed-initiative, let us consider the FIA
in more detail.

Fig.~\ref{fiaalgo} only reproduces the salient aspects of the FIA relevant
for our discussion. Compare the basic elements of the FIA to the stages in
Fig.~\ref{dialog-designs} (right). The Select phase corresponds to the
interpreter, the Collect phase gathers the user input, and
actions taken in the Process phase mimic the partial evaluator. Recall that 
`programs'
(scripts) in VoiceXML can be modeled by finite-state machines, hence
the mechanics of partial evaluation are considerably simplified and just
amount to filling the slot and removing it from further consideration.
Since the FIA repeatedly executes till there are no remaining form items,
the processing phase (Process) is effectively parameterized by the form-level
grammar file in Fig.~\ref{formgram}. In other words, the form-level grammar 
file not only enables slot filling, {\it it also implicitly directs the 
staging of interactions for mixed-initiative.} When the user
specifies `peperroni medium' in an utterace, not only does the grammar
file enable the recognition of the slots they correspond to (topping and size),
it also directs the FIA to simplify these slots (and remove them in
any subsequent interaction).

The form-level grammar file shown in Fig.~\ref{formgram} 
(which is also a specification of interaction staging) may make
VoiceXML's design appear overly complex. In reality, however,
we could have used the vanilla form-level 
grammar file in Fig.~\ref{formgram2}. While helping to 
realize mixed-initiative with Fig.~\ref{vpizza}, the new
form-level file (as does our model) also allows the possibility of 
utterances such as `pepperoni pepperoni,' or even, `pepperoni sausage!' 
Suitable semantics for such situations (including the role of 
side-effects) can be defined and accommodated in both the VoiceXML 
model and ours. It should thus be obvious to the reader that VoiceXML's 
dialog management architecture is actually implementing a mixed 
evaluation model (for conversational finite state machines), comprising 
interpretation and partial evaluation.

The VoiceXML specification~\cite{voicexml} refers to the form-level file 
as merely a `grammar file,' when it is actually also a specification of 
staging. Even though the grammar file serves the role of a language 
model in a voice application, we believe that
separating its two functionalities is important in understanding
mixed-initiative system design. 
A case in point is our study of personalizing 
interaction with web sites~\cite{pipe-tois}. There is no requirement for 
a `grammar file,' as there is usually no ambiguity about user clicks and 
typed-in keywords. In this context, the functionality provided by our model
is actually unmatched by any existing web-based interaction system (as
web interfaces are not typically designed for mixing initiative). A way to
incorporate mixed-initiative interaction into an existing interaction
at a web site is described in~\cite{pipe-tois}. 

\begin{table}
\centering
\begin{tabular}{|l|c|c|} \hline
\multicolumn{1}{|l|} {Software} &
\multicolumn{1}{c|} {Support for} &
\multicolumn{1}{c|} {Support for} \\ 
\multicolumn{1}{|l|} {Technology} &
\multicolumn{1}{c|} {Slot Simplification} &
\multicolumn{1}{c|} {Interaction Staging} \\ \hline
VoiceXML & $\surd$ & $\surd$ \\
Slot Filling Systems & $\surd$ & $\times$\\
Recognizer-Only APIs & $\times$ & $\times$ \\
\hline
\end{tabular}
\caption{Comparison of software technologies for voice-based
mixed-initiative applications.}
\label{compare-table}
\end{table}

\vspace{-0.1in}
\subsection{Other Implementation Technologies}
VoiceXML's FIA thus includes native support for slot filling, slot
simplification, and interaction staging. All of these are functions
enabled by partial evaluation in our model. Table~\ref{compare-table}
contrasts two other implementation approaches in terms of these aspects. 
In a purely slot-filling system, native support
is provided for simplifying slots from user utterances but extra code
needs to be written to model the control logic (for instance,
`the user still didn't specify his choice of size, so the question for
size should be repeated.'). Several commercial speech recognition vendors
provide APIs that operate at this level. In addition, many vendors support
low-level APIs that provide basic access to recognition results (i.e.,
text strings) but do not perform any additional processing. We refer
to these as recognizer-only APIs.
They serve more as raw 
speech recognition engines and require significant programming to first 
implement a slot-filling engine and, later, control logic to mimic all
possible opportunities for staging. Examples of the two latter technologies
can be seen in the commercial spoken dialog systems
market (from companies such as Nuance, IBM, and AT\&T). The study presented 
in this paper suggests a systematic way by which
their capabilities for mixed-initiative interaction can be assessed.

\vspace{-0.1in}
\section{Discussion}
\label{future}
Our work makes contributions to both partial evaluation and 
mixed-initiative interaction. For the partial evaluation community, we
have identified a novel application where the motivation is the staging 
of interaction (rather than speedup). Since programs (dialogs) are
used as specifications of interaction, they are {\it written to be
partially evaluated}; partial evaluation is hence not an `afterthought' 
or an optimization. A program can thus be thought of as a
compaction of all possible interaction sequences that involve mixing 
initiative. An interesting research issue is:
Given (i) a set of interaction sequences, and (ii) addressable information
(such as arguments and slot variables), determine (iii) the smallest program so
that every interaction sequence can be staged in the model 
of Fig.~\ref{dialog-designs} (right). This requires algorithms
to automatically decompose parts of interaction sequences into those
that are best addressed in the interpreter and those that can benefit from
representation and specialization by the partial evaluator. 

For mixed-initiative interaction, we have presented a 
programming model that accommodates all possibilities of staging, without 
explicit enumeration. The model makes a distinction between fixed-initiative
(and which has to be explicitly programmed) and mixed-initiative 
(specifications of which can be elegantly compressed for subsequent
partial evaluation). We have identified instantiations of this model in
VoiceXML and slot-filling APIs. We hope this observation will help
system designers gain additional insight into voice application design
strategies. 

It should be recalled that there are various facets of mixed-initiative 
that are not addressed in this paper. Extending our programming model to cover
these facets is an immediate direction of future research. For example, 
VoiceXML's design currently supports dialogs such as 
the following:

\begin{descit}{Dialog 5}
\vspace{-0.1in}
\begin{tabbing}
[x] \= abcdefab2 \= thiscanactuallybeamuchlongersentenceokay \kill
1 \> {\bf System:} \> Thank you for calling Joe's pizza ordering system.\\
2 \> {\bf System:} \> What size pizza would you like?\\
3 \> {\bf Caller 1:} \> What sizes do you have?\\
3 \> {\bf Caller 2:} \> Err.. Why don't you ask me the questions in
topping-crust-size order?\\
\end{tabbing}
\end{descit}
\vspace{-0.2in}

\noindent
{\it Caller 1}'s request, while demonstrating initiative, implies a dialog
with an optional stage (which cannot be modeled by partial
evaluation). Such a situation has to be trapped by the interpreter, not
by partial evaluation. {\it Caller 2} does specify a staging, but his 
staging poses constraints on the computer's initiative, not 
his own. Such a `meta-dialog' facet~\cite{mixed-hci} requires the 
ability to jump out 
of the current dialog; VoiceXML provides many elements for describing 
such transitions.

VoiceXML also provides certain `impure' features and side-effects in
its programming model. For instance, after selecting a size (say, medium),
the caller could retake the initiative in a different part of the dialog
and select a size again (this time, large). This will cause the new 
value to over-ride any existing value in the {\tt size} 
slot (see Fig.~\ref{fiaalgo}). In
our model, this implies the dynamic substitution of an earlier,
`evaluated out,' stage with a functional equivalent. Obviously, the dialog
manager has to maintain some state (across partial evaluations)
to accomplish this feature. We plan to investigate programming models suitable 
for these aspects. In addition, we plan to extend our software model 
beyond slot-and-filler structures, to include reasoning and exploiting 
context. 

Our long-term goal is to characterize mixed-initiative facets, not in
terms of initiative, interaction, or task models but in terms of the
opportunities for staging and the program transformation techniques that
can support such staging. This means that we can establish a 
taxonomy of mixed-initiative facets based on the transformation techniques
(e.g., partial evaluation, slicing) needed to realize them.
Such a taxonomy would also help connect the facets to design
models for interactive software systems.

\bibliographystyle{alpha}
\bibliography{pepm}

\end{document}

%% file: psfig-dvips.tex
\ifx\undefined\psfig\else \fi

\edef\psfigRestoreAt{\catcode`@=\number\catcode`@\relax}
\catcode`\@=11\relax
\newwrite\@unused
\def\typeout#1{{\let\protect\string\immediate\write\@unused{#1}}}
\def\figurepath{./}

\def\@nnil{\@nil}
\def\@empty{}
\def\@psdonoop#1\@@#2#3{}
\def\@psdo#1:=#2\do#3{\edef\@psdotmp{#2}\ifx\@psdotmp\@empty \else
    \expandafter\@psdoloop#2,\@nil,\@nil\@@#1{#3}\fi}
\def\@psdoloop#1,#2,#3\@@#4#5{\def#4{#1}\ifx #4\@nnil \else
       #5\def#4{#2}\ifx #4\@nnil \else#5\@ipsdoloop #3\@@#4{#5}\fi\fi}
\def\@ipsdoloop#1,#2\@@#3#4{\def#3{#1}\ifx #3\@nnil 
       \let\@nextwhile=\@psdonoop \else
      #4\relax\let\@nextwhile=\@ipsdoloop\fi\@nextwhile#2\@@#3{#4}}
\def\@tpsdo#1:=#2\do#3{\xdef\@psdotmp{#2}\ifx\@psdotmp\@empty \else
    \@tpsdoloop#2\@nil\@nil\@@#1{#3}\fi}
\def\@tpsdoloop#1#2\@@#3#4{\def#3{#1}\ifx #3\@nnil 
       \let\@nextwhile=\@psdonoop \else
      #4\relax\let\@nextwhile=\@tpsdoloop\fi\@nextwhile#2\@@#3{#4}}
\newread\ps@stream
\newif\ifnot@eof       
\newif\if@noisy        
\newif\if@atend        
\newif\if@psfile       
{\catcode`\%=12\global\gdef\epsf@start{
\def\epsf@PS{PS}
\def\epsf@getbb#1{%
\openin\ps@stream=#1
\ifeof\ps@stream\typeout{Error, File #1 not found}\else
   {\not@eoftrue \chardef\other=12
    \def\do##1{\catcode`##1=\other}\dospecials \catcode`\ =10
    \loop
       \if@psfile
	  \read\ps@stream to \epsf@fileline
       \else{
	  \obeyspaces
          \read\ps@stream to \epsf@tmp\global\let\epsf@fileline\epsf@tmp}
       \fi
       \ifeof\ps@stream\not@eoffalse\else
       \if@psfile\else
       \expandafter\epsf@test\epsf@fileline:. \\%
       \fi
          \expandafter\epsf@aux\epsf@fileline:. \\%
       \fi
   \ifnot@eof\repeat
   }\closein\ps@stream\fi}%
\long\def\epsf@test#1#2#3:#4\\{\def\epsf@testit{#1#2}
			\ifx\epsf@testit\epsf@start\else
\typeout{Warning! File does not start with `\epsf@start'.  It may not be a PostScript file.}
			\fi
			\@psfiletrue} 
{\catcode`\%=12\global\let\epsf@percent=
\long\def\epsf@aux#1#2:#3\\{\ifx#1\epsf@percent
   \def\epsf@testit{#2}\ifx\epsf@testit\epsf@bblit
	\@atendfalse
        \epsf@atend #3 . \\%
	\if@atend	
	   \if@verbose{
		\typeout{psfig: found `(atend)'; continuing search}
	   }\fi
        \else
        \epsf@grab #3 . . . \\%
        \not@eoffalse
        \global\no@bbfalse
        \fi
   \fi\fi}%
\def\epsf@grab #1 #2 #3 #4 #5\\{%
   \global\def\epsf@llx{#1}\ifx\epsf@llx\empty
      \epsf@grab #2 #3 #4 #5 .\\\else
   \global\def\epsf@lly{#2}%
   \global\def\epsf@urx{#3}\global\def\epsf@ury{#4}\fi}%
\def\epsf@atendlit{(atend)} 
\def\epsf@atend #1 #2 #3\\{%
   \def\epsf@tmp{#1}\ifx\epsf@tmp\empty
      \epsf@atend #2 #3 .\\\else
   \ifx\epsf@tmp\epsf@atendlit\@atendtrue\fi\fi}

\chardef\letter = 11
\chardef\other = 12

\newif \ifdebug 
\newif\ifc@mpute 
\c@mputetrue 

\let\then = \relax
\def\r@dian{pt }
\let\r@dians = \r@dian
\let\dimensionless@nit = \r@dian
\let\dimensionless@nits = \dimensionless@nit
\def\internal@nit{sp }
\let\internal@nits = \internal@nit
\newif\ifstillc@nverging
\def \Mess@ge #1{\ifdebug \then \message {#1} \fi}

{ 
	\catcode `\@ = \letter
	\gdef \nodimen {\expandafter \n@dimen \the \dimen}
	\gdef \term #1 #2 #3%
	       {\edef \t@ {\the #1}
		\edef \t@@ {\expandafter \n@dimen \the #2\r@dian}%
		\t@rm {\t@} {\t@@} {#3}%
	       }
	\gdef \t@rm #1 #2 #3%
	       {{%
		\count 0 = 0
		\dimen 0 = 1 \dimensionless@nit
		\dimen 2 = #2\relax
		\Mess@ge {Calculating term #1 of \nodimen 2}%
		\loop
		\ifnum	\count 0 < #1
		\then	\advance \count 0 by 1
			\Mess@ge {Iteration \the \count 0 \space}%
			\Multiply \dimen 0 by {\dimen 2}%
			\Mess@ge {After multiplication, term = \nodimen 0}%
			\Divide \dimen 0 by {\count 0}%
			\Mess@ge {After division, term = \nodimen 0}%
		\repeat
		\Mess@ge {Final value for term #1 of 
				\nodimen 2 \space is \nodimen 0}%
		\xdef \Term {#3 = \nodimen 0 \r@dians}%
		\aftergroup \Term
	       }}
	\catcode `\p = \other
	\catcode `\t = \other
	\gdef \n@dimen #1pt{#1} 
}

\def \Divide #1by #2{\divide #1 by #2} 

\def \Multiply #1by #2
       {{
	\count 0 = #1\relax
	\count 2 = #2\relax
	\count 4 = 65536
	\Mess@ge {Before scaling, count 0 = \the \count 0 \space and
			count 2 = \the \count 2}%
	\ifnum	\count 0 > 32767 
	\then	\divide \count 0 by 4
		\divide \count 4 by 4
	\else	\ifnum	\count 0 < -32767
		\then	\divide \count 0 by 4
			\divide \count 4 by 4
		\else
		\fi
	\fi
	\ifnum	\count 2 > 32767 
	\then	\divide \count 2 by 4
		\divide \count 4 by 4
	\else	\ifnum	\count 2 < -32767
		\then	\divide \count 2 by 4
			\divide \count 4 by 4
		\else
		\fi
	\fi
	\multiply \count 0 by \count 2
	\divide \count 0 by \count 4
	\xdef \product {#1 = \the \count 0 \internal@nits}%
	\aftergroup \product
       }}

\def\r@duce{\ifdim\dimen0 > 90\r@dian \then   
		\multiply\dimen0 by -1
		\advance\dimen0 by 180\r@dian
		\r@duce
	    \else \ifdim\dimen0 < -90\r@dian \then  
		\advance\dimen0 by 360\r@dian
		\r@duce
		\fi
	    \fi}

\def\Sine#1%
       {{%
	\dimen 0 = #1 \r@dian
	\r@duce
	\ifdim\dimen0 = -90\r@dian \then
	   \dimen4 = -1\r@dian
	   \c@mputefalse
	\fi
	\ifdim\dimen0 = 90\r@dian \then
	   \dimen4 = 1\r@dian
	   \c@mputefalse
	\fi
	\ifdim\dimen0 = 0\r@dian \then
	   \dimen4 = 0\r@dian
	   \c@mputefalse
	\fi
	\ifc@mpute \then
		\divide\dimen0 by 180
		\dimen0=3.141592654\dimen0
		\dimen 2 = 3.1415926535897963\r@dian 
		\divide\dimen 2 by 2 
		\Mess@ge {Sin: calculating Sin of \nodimen 0}%
		\count 0 = 1 
		\dimen 2 = 1 \r@dian 
		\dimen 4 = 0 \r@dian 
		\loop
			\ifnum	\dimen 2 = 0 
			\then	\stillc@nvergingfalse 
			\else	\stillc@nvergingtrue
			\fi
			\ifstillc@nverging 
			\then	\term {\count 0} {\dimen 0} {\dimen 2}%
				\advance \count 0 by 2
				\count 2 = \count 0
				\divide \count 2 by 2
				\ifodd	\count 2 
				\then	\advance \dimen 4 by \dimen 2
				\else	\advance \dimen 4 by -\dimen 2
				\fi
		\repeat
	\fi		
			\xdef \sine {\nodimen 4}%
       }}

\def\Cosine#1{\ifx\sine\UnDefined\edef\Savesine{\relax}\else
		             \edef\Savesine{\sine}\fi
	{\dimen0=#1\r@dian\multiply\dimen0 by -1
	 \advance\dimen0 by 90\r@dian
	 \Sine{\nodimen 0}
	 \xdef\cosine{\sine}
	 \xdef\sine{\Savesine}}}	      

\def\psdraft{
	\def\@psdraft{0}
}
\def\psfull{
	\def\@psdraft{100}
}

\psfull

\newif\if@draftbox
\def\psnodraftbox{
	\@draftboxfalse
}
\@draftboxtrue

\newif\if@prologfile
\newif\if@postlogfile
\def\pssilent{
	\@noisyfalse
}
\def\psnoisy{
	\@noisytrue
}
\psnoisy
\newif\if@bbllx
\newif\if@bblly
\newif\if@bburx
\newif\if@bbury
\newif\if@height
\newif\if@width
\newif\if@rheight
\newif\if@rwidth
\newif\if@angle
\newif\if@clip
\newif\if@verbose
\newif\if@scale
\def\@p@@sclip#1{\@cliptrue}

\def\@p@@sfile#1{\def\@p@sfile{null}%
	        \openin1=#1
		\ifeof1\closein1%
		       \openin1=\figurepath#1
			\ifeof1\typeout{Error, File #1 not found}
			   \if@bbllx\if@bblly\if@bburx\if@bbury
			      \def\@p@sfile{#1}%
			   \fi\fi\fi\fi
			\else\closein1
			    \edef\@p@sfile{\figurepath#1}%
                        \fi%
		 \else\closein1%
		       \def\@p@sfile{#1}%
		 \fi}
\def\@p@@sfigure#1{\def\@p@sfile{null}%
	        \openin1=#1
		\ifeof1\closein1%
		       \openin1=\figurepath#1
			\ifeof1\typeout{Error, File #1 not found}
			   \if@bbllx\if@bblly\if@bburx\if@bbury
			      \def\@p@sfile{#1}%
			   \fi\fi\fi\fi
			\else\closein1
			    \def\@p@sfile{\figurepath#1}%
                        \fi%
		 \else\closein1%
		       \def\@p@sfile{#1}%
		 \fi}

\def\@p@@sbbllx#1{
		\@bbllxtrue
		\dimen100=#1
		\edef\@p@sbbllx{\number\dimen100}
}
\def\@p@@sbblly#1{
		\@bbllytrue
		\dimen100=#1
		\edef\@p@sbblly{\number\dimen100}
}
\def\@p@@sbburx#1{
		\@bburxtrue
		\dimen100=#1
		\edef\@p@sbburx{\number\dimen100}
}
\def\@p@@sbbury#1{
		\@bburytrue
		\dimen100=#1
		\edef\@p@sbbury{\number\dimen100}
}
\def\@p@@sheight#1{
		\@heighttrue
		\dimen100=#1
   		\edef\@p@sheight{\number\dimen100}
}
\def\@p@@swidth#1{
		\@widthtrue
		\dimen100=#1
		\edef\@p@swidth{\number\dimen100}
}
\def\@p@@srheight#1{
		\@rheighttrue
		\dimen100=#1
		\edef\@p@srheight{\number\dimen100}
}
\def\@p@@srwidth#1{
		\@rwidthtrue
		\dimen100=#1
		\edef\@p@srwidth{\number\dimen100}
}
\def\@p@@sangle#1{
		\@angletrue
		\edef\@p@sangle{#1} 
}
\def\@p@@ssilent#1{ 
		\@verbosefalse
}
\def\@p@@sscale#1{
		\def\@p@scale{#1}
		\@scaletrue
}
\def\@p@@sprolog#1{\@prologfiletrue\def\@prologfileval{#1}}
\def\@p@@spostlog#1{\@postlogfiletrue\def\@postlogfileval{#1}}
\def\@cs@name#1{\csname #1\endcsname}
\def\@setparms#1=#2,{\@cs@name{@p@@s#1}{#2}}
\def\ps@init@parms{
		\@bbllxfalse \@bbllyfalse
		\@bburxfalse \@bburyfalse
		\@heightfalse \@widthfalse
		\@rheightfalse \@rwidthfalse
		\@scalefalse
		\def\@p@sbbllx{}\def\@p@sbblly{}
		\def\@p@sbburx{}\def\@p@sbbury{}
		\def\@p@sheight{}\def\@p@swidth{}
		\def\@p@srheight{}\def\@p@srwidth{}
		\def\@p@sangle{0}
		\def\@p@sfile{}
		\def\@p@scost{10}
		\def\@sc{}
		\@prologfilefalse
		\@postlogfilefalse
		\@clipfalse
		\if@noisy
			\@verbosetrue
		\else
			\@verbosefalse
		\fi
}
\def\parse@ps@parms#1{
	 	\@psdo\@psfiga:=#1\do
		   {\expandafter\@setparms\@psfiga,}}
\newif\ifno@bb
\def\bb@missing{
	\if@verbose{
		\typeout{psfig: searching \@p@sfile \space  for bounding box}
	}\fi
	\no@bbtrue
	\epsf@getbb{\@p@sfile}
        \ifno@bb \else \bb@cull\epsf@llx\epsf@lly\epsf@urx\epsf@ury\fi
}	
\def\bb@cull#1#2#3#4{
	\dimen100=#1 bp\edef\@p@sbbllx{\number\dimen100}
	\dimen100=#2 bp\edef\@p@sbblly{\number\dimen100}
	\dimen100=#3 bp\edef\@p@sbburx{\number\dimen100}
	\dimen100=#4 bp\edef\@p@sbbury{\number\dimen100}
	\no@bbfalse
}

\newdimen\p@intvaluex
\newdimen\p@intvaluey
\newdimen\@ffsetvalue
\newdimen\x@ffsetvalue
\newdimen\y@ffsetvalue

\def\compute@offset#1#2{{\dimen0=#1 sp\dimen1=#2 sp
			\advance\dimen1 by -\dimen0
			\dimen1=\sine\dimen1
			\dimen0=\cosine\dimen1
			\ifdim\dimen0<0sp \dimen1=0sp \fi
			\global\@ffsetvalue=\dimen1}}

\def\rotate@#1#2{{\dimen0=#1 sp\dimen1=#2 sp
		  \global\p@intvaluex=\cosine\dimen0
		  \dimen3=\sine\dimen1
		  \global\advance\p@intvaluex by -\dimen3
		  \global\p@intvaluey=\sine\dimen0
		  \dimen3=\cosine\dimen1
		  \global\advance\p@intvaluey by \dimen3
		  }}
\def\compute@bb{
		\no@bbfalse
		\if@bbllx \else \no@bbtrue \fi
		\if@bblly \else \no@bbtrue \fi
		\if@bburx \else \no@bbtrue \fi
		\if@bbury \else \no@bbtrue \fi
		\ifno@bb \bb@missing \fi
		\ifno@bb \typeout{FATAL ERROR: no bb supplied or found}
			\no-bb-error
		\fi
		\if@angle 
			\Sine{\@p@sangle}\Cosine{\@p@sangle}
			\compute@offset{\@p@sbblly}{\@p@sbbury}
			\x@ffsetvalue=\@ffsetvalue
			\compute@offset{\@p@sbburx}{\@p@sbbllx}
			\y@ffsetvalue=\@ffsetvalue

			\rotate@{\@p@sbbllx}{\@p@sbblly}
			\advance\p@intvaluex by -\x@ffsetvalue
			\advance\p@intvaluey by -\y@ffsetvalue
			\edef\@p@sbbllx{\number\p@intvaluex}
			\edef\@p@sbblly{\number\p@intvaluey}

			\rotate@{\@p@sbburx}{\@p@sbbury}
			\advance\p@intvaluex by \x@ffsetvalue
			\advance\p@intvaluey by \y@ffsetvalue
			\edef\@p@sbburx{\number\p@intvaluex}
			\edef\@p@sbbury{\number\p@intvaluey}
			{
			 \count0=\@p@sbbllx \count1=\@p@sbblly
		 	 \count2=\@p@sbburx \count3=\@p@sbbury
			 \dimen0=\@p@sbbllx sp\dimen1=\@p@sbblly sp
		 	 \dimen2=\@p@sbburx sp\dimen3=\@p@sbbury sp
			 \dimen203=\dimen2 \advance\dimen203 by -\dimen0
			 \dimen204=\dimen3 \advance\dimen204 by -\dimen1
			 \ifdim\dimen203<0sp 
			      \count203=\count2 \count2=\count0 
			      \count0=\count203 
			      \global\edef\@p@sbbllx{\number\count0}
			      \global\edef\@p@sbburx{\number\count2}
			 \fi
			 \ifdim\dimen204<0sp 
			       \count204=\count3
			       \count3=\count1
			       \count1=\count204
			       \global\edef\@p@sbblly{\number\count1}
			       \global\edef\@p@sbbury{\number\count3}
			 \fi
			}
		\fi
		\count203=\@p@sbburx
		\count204=\@p@sbbury
		\advance\count203 by -\@p@sbbllx
		\advance\count204 by -\@p@sbblly
		\edef\@bbw{\number\count203}
		\edef\@bbh{\number\count204}
}
\def\in@hundreds#1#2#3{\count240=#2 \count241=#3
		     \count100=\count240	
		     \divide\count100 by \count241
		     \count101=\count100
		     \multiply\count101 by \count241
		     \advance\count240 by -\count101
		     \multiply\count240 by 10
		     \count101=\count240	
		     \divide\count101 by \count241
		     \count102=\count101
		     \multiply\count102 by \count241
		     \advance\count240 by -\count102
		     \multiply\count240 by 10
		     \count102=\count240	
		     \divide\count102 by \count241
		     \count200=#1\count205=0
		     \count201=\count200
			\multiply\count201 by \count100
		 	\advance\count205 by \count201
		     \count201=\count200
			\divide\count201 by 10
			\multiply\count201 by \count101
			\advance\count205 by \count201
		     \count201=\count200
			\divide\count201 by 100
			\multiply\count201 by \count102
			\advance\count205 by \count201
		     \edef\@result{\number\count205}
}
\def\@ScaleInHundreds#1{
		\in@hundreds{#1}{\@p@scale}{100}
		\edef#1{\@result}
}
\def\compute@wfromh{
		\in@hundreds{\@p@sheight}{\@bbw}{\@bbh}
		\edef\@p@swidth{\@result}
}
\def\compute@hfromw{
		\in@hundreds{\@p@swidth}{\@bbh}{\@bbw}
		\edef\@p@sheight{\@result}
}
\def\compute@handw{
		\if@height 
			\if@width
			\else
				\compute@wfromh
			\fi
		\else 
			\if@width
				\compute@hfromw
			\else
				\edef\@p@sheight{\@bbh}
				\edef\@p@swidth{\@bbw}
			\fi
		\fi
}
\def\compute@resv{
		\if@rheight \else \edef\@p@srheight{\@p@sheight} \fi
		\if@rwidth \else \edef\@p@srwidth{\@p@swidth} \fi
}
\def\compute@sizes{
	\compute@bb
	\compute@handw
	\compute@resv
}
\def\psfig#1{\vbox {
	\ps@init@parms
	\parse@ps@parms{#1}
	\compute@sizes
	\if@scale
                \if@verbose
                        \typeout{psfig: scaling by \@p@scale}
                \fi
                \@ScaleInHundreds{\@p@swidth}
                \@ScaleInHundreds{\@p@sheight}
                \@ScaleInHundreds{\@p@srwidth}
                \@ScaleInHundreds{\@p@srheight}
        \fi
	\ifnum\@p@scost<\@psdraft{
		\if@verbose{
			\typeout{psfig: including \@p@sfile \space }
		}\fi
		\special{ps::[begin] 	\@p@swidth \space \@p@sheight \space
				\@p@sbbllx \space \@p@sbblly \space
				\@p@sbburx \space \@p@sbbury \space
				startTexFig \space }
		\if@angle
			\special {ps:: \@p@sangle \space rotate \space} 
		\fi
		\if@clip{
			\if@verbose{
				\typeout{(clip)}
			}\fi
			\special{ps:: doclip \space }
		}\fi
		\if@prologfile
		    \special{ps: plotfile \@prologfileval \space } \fi
		\special{ps: plotfile \@p@sfile \space }
		\if@postlogfile
		    \special{ps: plotfile \@postlogfileval \space } \fi
		\special{ps::[end] endTexFig \space }
		\vbox to \@p@srheight true sp{
			\hbox to \@p@srwidth true sp{
				\hss
			}
		\vss
		}
	}\else{
		\if@draftbox{		
			\hbox{\fbox{\vbox to \@p@srheight true sp{
			\vss
			\hbox to \@p@srwidth true sp{ \hss \@p@sfile \hss }
			\vss
			}}}
		}\else{
			\vbox to \@p@srheight true sp{
			\vss
			\hbox to \@p@srwidth true sp{\hss}
			\vss
			}
		}\fi

	}\fi
}}
\def\psglobal{\typeout{psfig: PSGLOBAL is OBSOLETE; use psprint -m instead}}
\psfigRestoreAt